\documentclass[11pt,twoside]{article}
\usepackage{graphicx}
\usepackage{subfigure}
\usepackage{fancyheadings}
\usepackage{amsmath}
\usepackage{amssymb}
\usepackage{cite}
\usepackage{color,colortbl}

\definecolor{darkishgreen}{RGB}{39,203,22}
\definecolor{LightCyan}{rgb}{0.88,1,1}
\definecolor{Gray}{gray}{0.9}
\definecolor{lightRed}{RGB}{230,170,150}
\definecolor{modRed}{RGB}{230,82,90}
\definecolor{strongRed}{RGB}{230,6,6}

\begin{document}
\newcommand{\pst}{\hspace*{1.5em}}

\newcommand{\rigmark}{\em Journal of Russian Laser Research}
\newcommand{\lemark}{\em Volume 30, Number 5, 2009}

%\lhead[\fancyplain{\rigmark, {\em \lemark}}{\rigmark}]{\fancyplain{\rigmark, {\em \lemark}}{\lemark}}
%\chead{}\rhead[\fancyplain{}{\lemark}]{\fancyplain{}{\rigmark}}
%\plainfootrulewidth 0.4pt
\newcommand{\be}{\begin{equation}}
\newcommand{\ee}{\end{equation}}
\newcommand{\bm}{\boldmath}
\newcommand{\ds}{\displaystyle}
\newcommand{\bea}{\begin{eqnarray}}
\newcommand{\eea}{\end{eqnarray}}
\newcommand{\ba}{\begin{array}}
\newcommand{\ea}{\end{array}}
\newcommand{\arcsinh}{\mathop{\rm arcsinh}\nolimits}
\newcommand{\arctanh}{\mathop{\rm arctanh}\nolimits}
\newcommand{\bc}{\begin{center}}
\newcommand{\ec}{\end{center}}

\thispagestyle{plain}

\label{sh}

%\lfoot[\fancyplain{\ \\[1mm] \thepage}{\ \\[1mm]\thepage}]{\fancyplain{}{}}

\begin{center} {\Large \bf
\begin{tabular}{c}
ENTROPIC INEQUALITIES AND PROPERTIES
\\[-1mm]
OF SOME SPECIAL FUNCTIONS
\end{tabular}
 } \end{center}

\bigskip

\bigskip

\begin{center} {\bf
V.I. Man'ko$^{1}$ and L.A. Markovich$^{2*}$
}\end{center}

\medskip

\begin{center}
{\it
$^1$P.N. Lebedev Physical Institute, Russian Academy of Sciences\\
Leninskii Prospect 53, Moscow 119991, Russia

\smallskip

$^2$Institute of Control Sciences, Russian Academy of Sciences\\
Profsoyuznaya 65, Moscow 117997, Russia
}
\smallskip

$^*$Corresponding author e-mail:~~~kimo1~@~mail.ru\\
\end{center}

\begin{abstract}\noindent
Using known entropic and information inequalities new inequalities for some classical polynomials
are obtained. Examples of Jacobi and Legendre polynomials are considered.
\end{abstract}

\medskip


\section{Introduction}
\pst
It is known \cite{Vil} that there exists a possibility to get some relations for special functions which turn out
to be the matrix elements of irreducible unitary representation of compact and noncompact groups.
\par On the other hand, in classical probability theory  and in quantum tomographic approach \cite{Ibort}
for description of quantum states the specific probability distributions expressed in terms of some special functions appear
in a natural way. Since entropies  are determined by the probability distributions and there exist relations,
in particular, in the form of entropic and information inequalities, one can apply these   inequalities to get
some new inequalities for the special functions.
\par For one random variable the probability distribution which appears as a result of experiment with finite number of outcomes is characterized by
the Shannon entropy \cite{Shannon}. The results of experiments where two random variables are measured can be associated with a joint probability distribution. The
distribution is connected with $N=N_1\cdot N_2$ outcomes where for the first random variable we have $N_1$ results and for second random variable we have $N_2$ results.
The joint probability distribution and dependence between two random variables determine two marginal probability distributions by the Sklar's theorem \cite{Nelsen}. For these  three probability distributions one can calculate Shannon entropies \cite{Shannon}.
These entropies satisfy the inequality called the subadditivity condition \cite{Lieb}. The entropic inequalities for the bipartite systems were used in \cite{Chernega,Mendes} in the
framework of the tomographic probability representation of quantum mechanics to characterize two degrees of quantum correlations in the systems. On the other hand the
mathematical structure of the subadditivity condition permits to apply   this inequality in all cases where the set of nonnegative numbers or functions appears and
the sum of numbers or functions equals to unity.
\par The aim of our work is to consider the unitary matrices connected with the irreducible representation of the rotation group and other groups and construct probability
distributions creating from
entropic inequalities the inequalities for such special functions  as Jacobi and Legendre polynomials.
\section{Inequalities for $SU(2)$ - representation matrix elements}
\pst
It is known that the unitary irreducible representations of rotation group with spins (or $SU(2)$)   are expressed in terms of Jacobi polynomials \cite{Vil,Landau}.
The squared modules of the matrix elements are
\begin{eqnarray}\label{1}
\Big|d_{m',m}^{(j)}(\beta)\Big|^2\!\!\!\!\!&=&\!\!\!\!\frac{(j+m')!(j-m')!}{(j+m)!(j-m)!}\left(\cos\left(\frac{\beta}{2}\right)^{m'+m}\sin\left(\frac{\beta}{2}\right)^{m'-m}
P_{j-m'}^{(m'-m,m'+m)}(\cos\beta)\right)^2,
\end{eqnarray}
where $P_{j-m'}^{(m'-m,m'+m)}(\cos\beta)$ denote the Jacobi polynomials \cite{Landau}
\begin{eqnarray*}
P_{n}^{(a,b)}(z)&=& \frac{(-1)^n}{2^nn!}(1-z)^{-a}(1+z)^{-b}\frac{d^n}{dz^n}(1-z)^{a+n}(1+z)^{b+n}.
\end{eqnarray*}
The following relation
\begin{eqnarray}\label{11}
d_{m',m}^{(j)}(\beta)&=&(-1)^{m'-m}d_{m,m'}^{(j)}(\beta)=d_{-m,-m'}^{(j)}(\beta)
\end{eqnarray}
holds.
We shall apply the generic inequalities for probabilities expressed in terms of Shannon entropies to these matrix elements.
The point is that one has $\Big|d_{m'm}^{(j)}(\beta)\Big|^2\geq0$ and
\begin{eqnarray}\label{2}
\sum\limits_{m'=-j}^{j}\Big|d_{m',m}^{(j)}(\beta)\Big|^2&=&1,\quad
\sum\limits_{m=-j}^{j}\Big|d_{m',m}^{(j)}(\beta)\Big|^2=1.
\end{eqnarray}
Thus the values $\Big|d_{m'm}^{(j)}(\beta)\Big|^2$ can be considered as probabilities.
We denote these probabilities as
\begin{eqnarray}\label{3}
P_{m'}^{(j)}(\beta)&=&\Big|d_{m',m}^{(j)}(\beta)\Big|^2,\quad P_{m}^{(j)}(\beta)=\Big|d_{m',m}^{(j)}(\beta)\Big|^2.
\end{eqnarray}
We will use the map of the numbers $m'$ and $m$ onto the numbers $1,2,\ldots,N=2j+1$ using the following rule
\begin{eqnarray*}
-j\Rightarrow 1,\quad -j+1\Rightarrow 2,\quad\ldots,\quad j \Rightarrow N.
\end{eqnarray*}
Thus we can study the relation which can be obtained by considering the probability vector $\overrightarrow{p}=(p_1,p_2,\ldots,p_N)$, where
$\sum\limits_{k=1}^{N}p_k=1$, $p_k\geq0$ hold.
\par The Shannon entropy associated with the probability vector $\overrightarrow{p}$ is determined by
\begin{eqnarray}\label{10}H_{p}&=&-\sum\limits_{k}p_{k}\ln p_{k}.
\end{eqnarray}

\section{Example of the spin $j=3/2$}
\pst
Let us discuss the arbitrary probability distribution identified with  the $4$-vector $\overrightarrow{p}=(p_1,p_2,p_3,p_4)$.
For the probability vector $\overrightarrow{p}$ we can introduce the following invertible map of the indices
\begin{eqnarray}\label{14}
1\Leftrightarrow 11,\quad 2\Leftrightarrow 12,\quad 3 \Leftrightarrow 21,\quad 4 \Leftrightarrow 22.
\end{eqnarray}
Then the probabilities are given in the form of the matrix $(p_{il})$ which could be interpreted as a joint probability of the bipartite system
(two coins)
\begin{eqnarray*}
p_1\Leftrightarrow p_{11},\quad p_2\Leftrightarrow p_{12},\quad p_3 \Leftrightarrow p_{21},\quad p_4 \Leftrightarrow p_{22}.
\end{eqnarray*}
It means that we mapped the probabilities for the system without subsystems into joint probability distribution associated with the bipartite system. Marginal probability distributions determined by the joint probability distribution are
\begin{eqnarray}\label{15}\pi_i&=&\sum\limits_{l=1}^2p_{il}=p_{i1}+p_{i2},\\ \nonumber
\Pi_l&=&\sum\limits_{i=1}^2p_{il}=p_{1l}+p_{2l}.
\end{eqnarray}
Thus, according to definition of Shannon entropy \eqref{10} the entropies associated with the initial probability distribution and two marginals are
\begin{eqnarray}\label{6}H_{p}&=&-p_1\ln p_1 -p_2\ln p_2-p_3\ln p_3-p_4\ln p_4,\\\nonumber
H_{\pi}&=&-(p_1+p_2)\ln(p_1+p_2)-(p_3+p_4)\ln(p_3+p_4),\\\nonumber
H_{\Pi}&=&-(p_1+p_3)\ln(p_1+p_3)-(p_2+p_4)\ln(p_2+p_4).
\end{eqnarray}
It is known that the Shannon entropies associated with the bipartite system satisfy some inequalities.
The following inequality which is called the subadditivity condition reads
\begin{eqnarray*}&&H_{\pi}+H_{\Pi}\geq H_{p}.
\end{eqnarray*}
The Shannon information is defined as the difference of the sum of the entropies of the subsystems and entropy of the bipartite system, i.e.
\begin{eqnarray}\label{7}I(\beta)&=&H_{\pi}(\beta)+H_{\Pi}(\beta)-H_{p}(\beta).
\end{eqnarray}
Obviously the Shannon information satisfies the inequality $I(\beta)\geq 0$ for all angles $\beta$.
\par Now we focus on the particular probability distribution determined by matrix elements of four dimensional irreducible representation of the group $SU(2)$ which corresponds to spin $j=3/2$.
\par
For $j=3/2$ the numbers $m$ and $m'$ which are spin projections on $z$-axis take the values  $-3/2$, $-1/2$, $1/2$, $3/2$. All elements of \eqref{1}
are represented in Table \ref{tab:1},
%\cellcolor{Gray3}
\begin{table}[Ht]
    \begin{center}
    \begin{tabular}{c|c|c|c|c}
    $m'\setminus m$ &$3/2$ & $1/2$ &$-1/2$ & $-3/2$ \\
    \hline
    $3/2$  &$\widetilde{d}_{3/2,3/2}^{(j)}(\beta)$&$\widetilde{d}_{3/2,1/2}^{(j)}(\beta)$ &$\widetilde{d}_{3/2,-1/2}^{(j)}(\beta)$&$\widetilde{d}_{3/2,-3/2}^{(j)}(\beta)$\\
    \hline
     $1/2$&$\widetilde{d}_{1/2,3/2}^{(j)}(\beta)$&$\widetilde{d}_{1/2,1/2}^{(j)}(\beta)$&$\widetilde{d}_{1/2,-1/2}^{(j)}(\beta)$&$\widetilde{d}_{1/2,-3/2}^{(j)}(\beta)$\\
    \hline
    $-1/2$  &$\widetilde{d}_{-1/2,3/2}^{(j)}(\beta)$&$\widetilde{d}_{-1/2,1/2}^{(j)}(\beta)$&$\widetilde{d}_{-1/2,-1/2}^{(j)}(\beta)$&$\widetilde{d}_{-1/2,-3/2}^{(j)}(\beta)$\\
    \hline
     $-3/2$&$\widetilde{d}_{-3/2,3/2}^{(j)}(\beta)$&$\widetilde{d}_{-3/2,1/2}^{(j)}(\beta)$&$\widetilde{d}_{-3/2,-1/2}^{(j)}(\beta)$&$\widetilde{d}_{-3/2,-3/2}^{(j)}(\beta)$\\
    \end{tabular}
\caption{\label{tab:1}}
    \end{center}
    \end{table}
    where it was used the notation $\Big|d_{m',m}^{(j)}(\beta)\Big|^2\equiv\widetilde{d}_{m',m}^{(j)}(\beta)$.
\par Using \eqref{11} we can write the following relation for $\widetilde{d}_{m',m}^{(j)}(\beta)$
\begin{eqnarray*}
\widetilde{d}_{m',m}^{(j)}(\beta)&=&\widetilde{d}_{m,m'}^{(j)}(\beta)=\widetilde{d}_{-m,-m'}^{(j)}(\beta).
\end{eqnarray*}
Therefore, it is also straightforward to verify the following equations for the matrix elements
\begin{eqnarray}\label{12}\widetilde{d}_{3/2,3/2}^{(j)}(\beta)&=&\widetilde{d}_{-3/2,-3/2}^{(j)}(\beta),\quad
\widetilde{d}_{-3/2,3/2}^{(j)}(\beta)=\widetilde{d}_{3/2,-3/2}^{(j)}(\beta),\\\nonumber
\widetilde{d}_{1/2,1/2}^{(j)}(\beta)&=&\widetilde{d}_{-1/2,-1/2}^{(j)}(\beta),\quad
\widetilde{d}_{-1/2,1/2}^{(j)}(\beta)=\widetilde{d}_{1/2,-1/2}^{(j)}(\beta),\\\nonumber
\widetilde{d}_{3/2,1/2}^{(j)}(\beta)&=&\widetilde{d}_{1/2,3/2}^{(j)}(\beta)=
\widetilde{d}_{-3/2,-1/2}^{(j)}(\beta)=\widetilde{d}_{-1/2,-3/2}^{(j)}(\beta),\\\nonumber
\widetilde{d}_{3/2,-1/2}^{(j)}(\beta)&=&\widetilde{d}_{-1/2,3/2}^{(j)}(\beta)=
\widetilde{d}_{3/2,-1/2}^{(j)}(\beta)=\widetilde{d}_{1/2,-3/2}^{(j)}(\beta).
\end{eqnarray}
Hence  probabilities associated with them are also equal.
\par The probability vector $\overrightarrow{p}$ can be chosen as in \eqref{2}. For example, for the fixed $m=3/2$ and considering all elements with different $m'$, i.e. the elements of the first column in the Table \ref{tab:1} we obtain
\begin{eqnarray}\label{4}p_1(\beta)&=&\widetilde{d}_{3/2,3/2}^{(j)}(\beta)=\cos\left(\beta/2\right)^6P_{0}^{(0,3)}(\cos\beta)^2
=\frac{(\cos\beta+1)^3}{8},\\\nonumber
 p_2(\beta)&=&\widetilde{d}_{1/2,3/2}^{(j)}(\beta)=\frac{\cos\left(\beta/2\right)^4}{3\sin\left(\beta/2\right)^2}
 P_{1}^{(-1,2)}(\cos\beta)^2
= 3\sin\left(\beta/2\right)^2\left(\sin\left(\beta/2\right)^2-1\right)^2,\\\nonumber
p_3(\beta)&=&\widetilde{d}_{-1/2,3/2}^{(j)}(\beta)=\frac{\cos\left(\beta/2\right)^2}{3\sin\left(\beta/2\right)^4}P_{2}^{(-2,1)}(\cos\beta)^2
=\frac{3}{8}\left(\cos\beta-1\right)^2(\cos\beta+1),\\\nonumber
 p_4(\beta)&=&\widetilde{d}_{-3/2,3/2}^{(j)}(\beta)=\frac{1}{\sin\left(\beta/2\right)^6}P_{3}^{(-3,0)}(\cos\beta)^2
= -\frac{(\cos\beta-1)^3}{8}.
\end{eqnarray}
%cot- ctg
The sum of latter probabilities is equal to $\sum\limits_{k}p_{k}(\beta)=1$ for any angle $\beta$.
\par For example, for the angle $\beta=1$ the probabilities are the following
\begin{eqnarray*}&&p_1(1)=0.4568019,\quad p_2(1)=0.40899267,\quad
p_3(1)=0.1220624,\quad p_4(1)=0.012143
\end{eqnarray*}
and their sum is equal to $1$.
\par It is possible to fix $m'$ and consider all matrix elements of the $m'$th row of the Table \ref{tab:1}. Note that it is possible to construct vector   $\overrightarrow{p}$ by a variety of other combinations of elements in the Table \ref{tab:1} . This conclusion is a direct consequence of the equality \eqref{12}. It is only necessary to remember that the sum of the elements of such a vector must always be equal to one.
\par This gives us the opportunity to form a large number of inequalities based on inequality  \eqref{7} for the Shannon information.
\par
\begin{figure}[ht]
\begin{center}
\begin{minipage}[ht]{0.49\linewidth}
\includegraphics[width=1\linewidth]{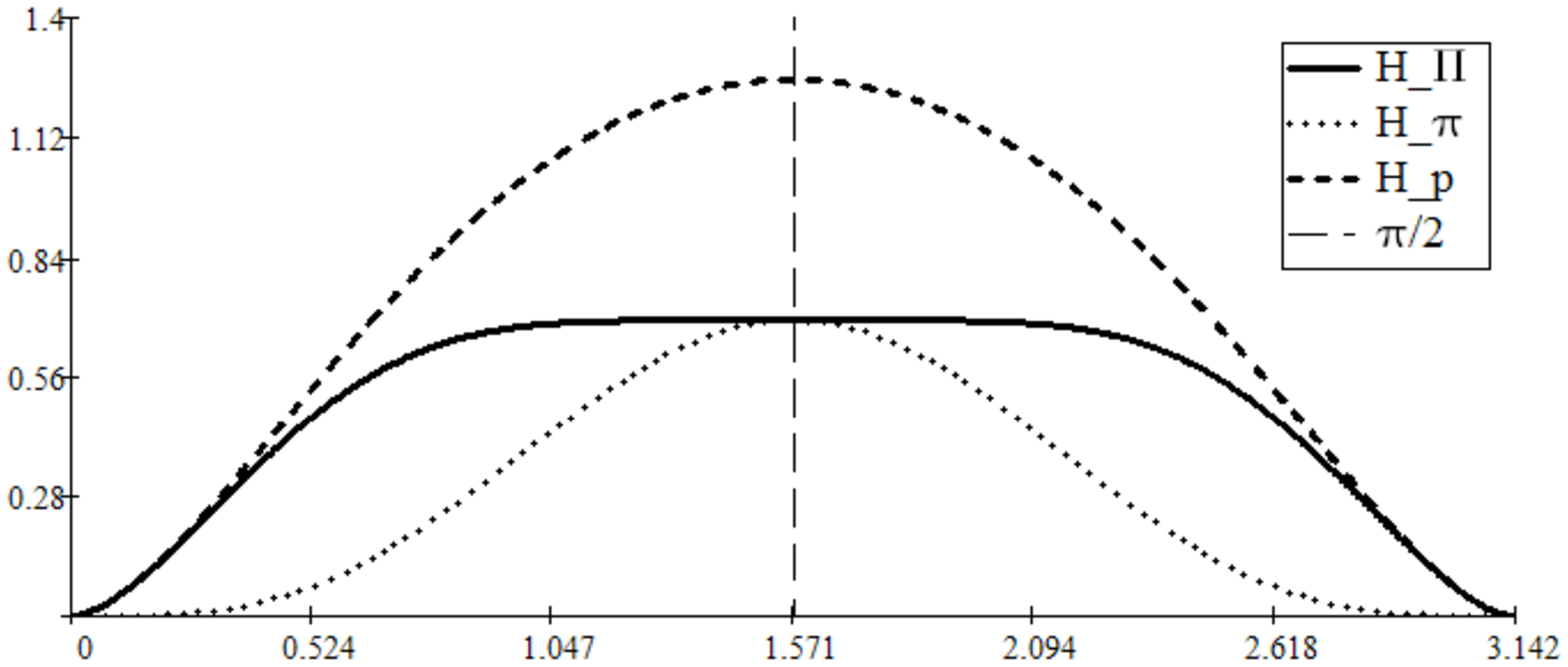}
\vspace{-4mm}
\caption{Entropies $H_{\Pi}(\beta)$, $H_{\pi}(\beta)$, $H_{p}(\beta)$ for probabilities \eqref{4}}
\label{fig:1}
\end{minipage}
\hfill
\begin{minipage}[ht]{0.49\linewidth}
\includegraphics[width=1\linewidth]{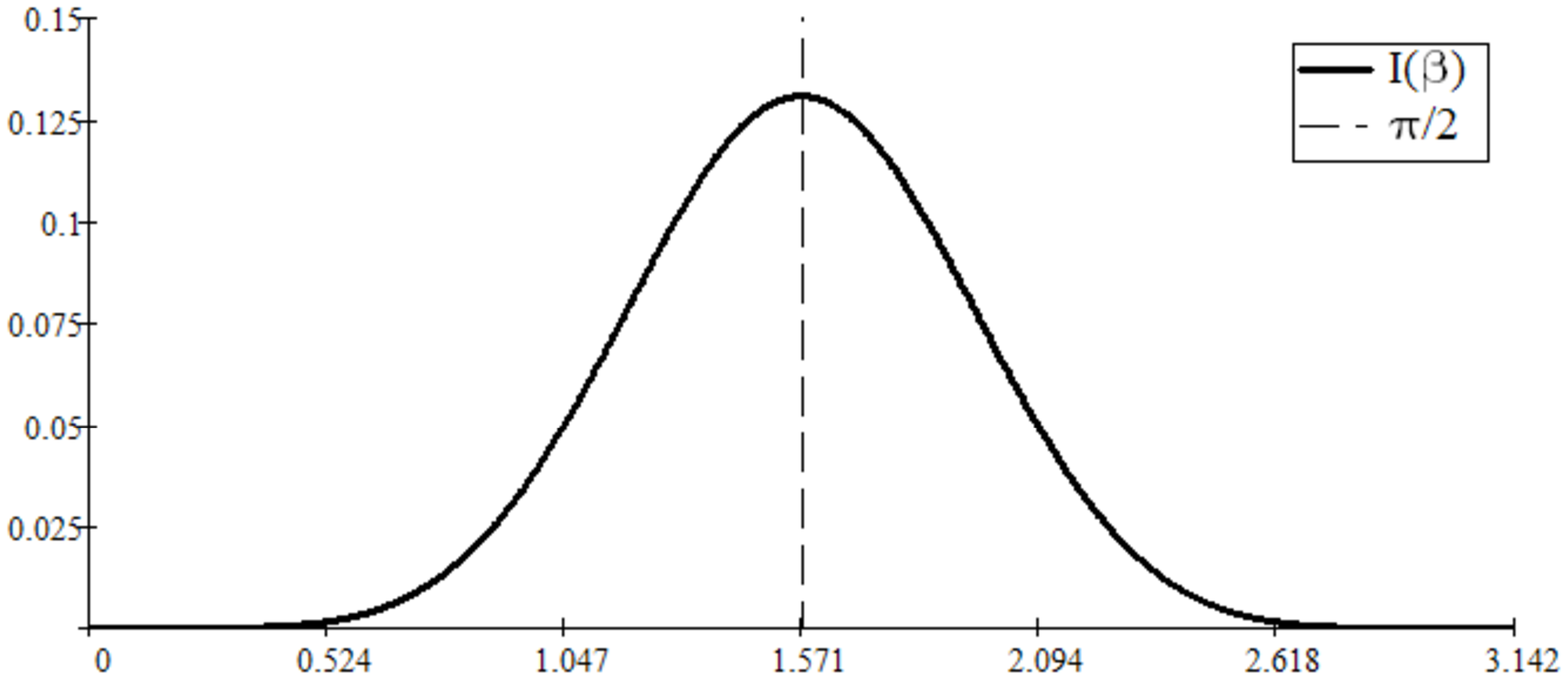}
\vspace{-4mm}
\caption{Information $I(\beta)$ for probabilities \eqref{4}}
\label{fig:2}
\end{minipage}
\end{center}
\end{figure}

 Let us introduce the following notation $\left(P_{j-m'}^{(m'-m,m'+m)}(\cos\beta)\right)^2\equiv \widetilde{P}_{j-m'}^{(m'-m,m'+m)}$. Substituting \eqref{4} in \eqref{6} and \eqref{7} we can write the following inequality
\begin{eqnarray}\label{17}&&-\left(\frac{\widetilde{P}_{3}^{(-3,0)}}{\sin\left(\beta/2\right)^6}+\frac{\widetilde{P}_{2}^{(-2,1)}\cos\left(\beta/2\right)^2}{3\sin\left(\beta/2\right)^4}\right)
\ln\left(\frac{\widetilde{P}_{3}^{(-3,0)}}{\sin\left(\beta/2\right)^6}+\frac{\widetilde{P}_{2}^{(-2,1)}\cos\left(\beta/2\right)^2}{3\sin\left(\beta/2\right)^4}\right)\\\nonumber
&-&\left(\cos\left(\beta/2\right)^6\widetilde{P}_{0}^{(0,3)}+\frac{\cos\left(\beta/2\right)^4}{3\sin\left(\beta/2\right)^2}
 \widetilde{P}_{1}^{(-1,2)}\right)
\ln\left(\cos\left(\beta/2\right)^6\widetilde{P}_{0}^{(0,3)}+\frac{\cos\left(\beta/2\right)^4}{3\sin\left(\beta/2\right)^2}
 \widetilde{P}_{1}^{(-1,2)}\right)\\\nonumber
&-&\left(\cos\left(\beta/2\right)^6\widetilde{P}_{0}^{(0,3)}+\frac{\widetilde{P}_{2}^{(-2,1)}\cos\left(\beta/2\right)^2}{3\sin\left(\beta/2\right)^4}\right)
\ln\left(\cos\left(\beta/2\right)^6\widetilde{P}_{0}^{(0,3)}+\frac{\widetilde{P}_{2}^{(-2,1)}\cos\left(\beta/2\right)^2}{3\sin\left(\beta/2\right)^4}\right)\\\nonumber
&-&\left(\frac{\widetilde{P}_{3}^{(-3,0)}}{\sin\left(\beta/2\right)^6}+\frac{\cos\left(\beta/2\right)^4}{3\sin\left(\beta/2\right)^2}
 \widetilde{P}_{1}^{(-1,2)}\right)
\ln\left(\frac{\widetilde{P}_{3}^{(-3,0)}}{\sin\left(\beta/2\right)^6}+\frac{\cos\left(\beta/2\right)^4}{3\sin\left(\beta/2\right)^2}
 \widetilde{P}_{1}^{(-1,2)}\right)\\\nonumber
&+&\cos\left(\beta/2\right)^6\widetilde{P}_{0}^{(0,3)}
\ln\left(\cos\left(\beta/2\right)^6\widetilde{P}_{0}^{(0,3)}\right)
+\frac{\cos\left(\beta/2\right)^4}{3\sin\left(\beta/2\right)^2}
 \widetilde{P}_{1}^{(-1,2)}
\ln\left(\frac{\cos\left(\beta/2\right)^4}{3\sin\left(\beta/2\right)^2}
 \widetilde{P}_{1}^{(-1,2)}\right)\\\nonumber
&+&\frac{\widetilde{P}_{2}^{(-2,1)}\cos\left(\beta/2\right)^2}{3\sin\left(\beta/2\right)^4}
\ln\left(\frac{\widetilde{P}_{2}^{(-2,1)}\cos\left(\beta/2\right)^2}{3\sin\left(\beta/2\right)^4}\right)
+\frac{\widetilde{P}_{3}^{(-3,0)}}{\sin\left(\beta/2\right)^6}
\ln\left(\frac{\widetilde{P}_{3}^{(-3,0)}}{\sin\left(\beta/2\right)^6}\right)
\geq0
\end{eqnarray}
Entropies $H_{\Pi}(\beta)$, $H_{\pi}(\beta)$, $H_{p}(\beta)$ and the Shannon information $I(\beta)$ for probabilities \eqref{4} are shown in Figures \ref{fig:1} and \ref{fig:2}. Obviously the information reaches its maximum value at the point $\beta=\pi/2$.
\par
\begin{figure}[ht]
\begin{center}
\begin{minipage}[ht]{0.49\linewidth}
\includegraphics[width=1\linewidth]{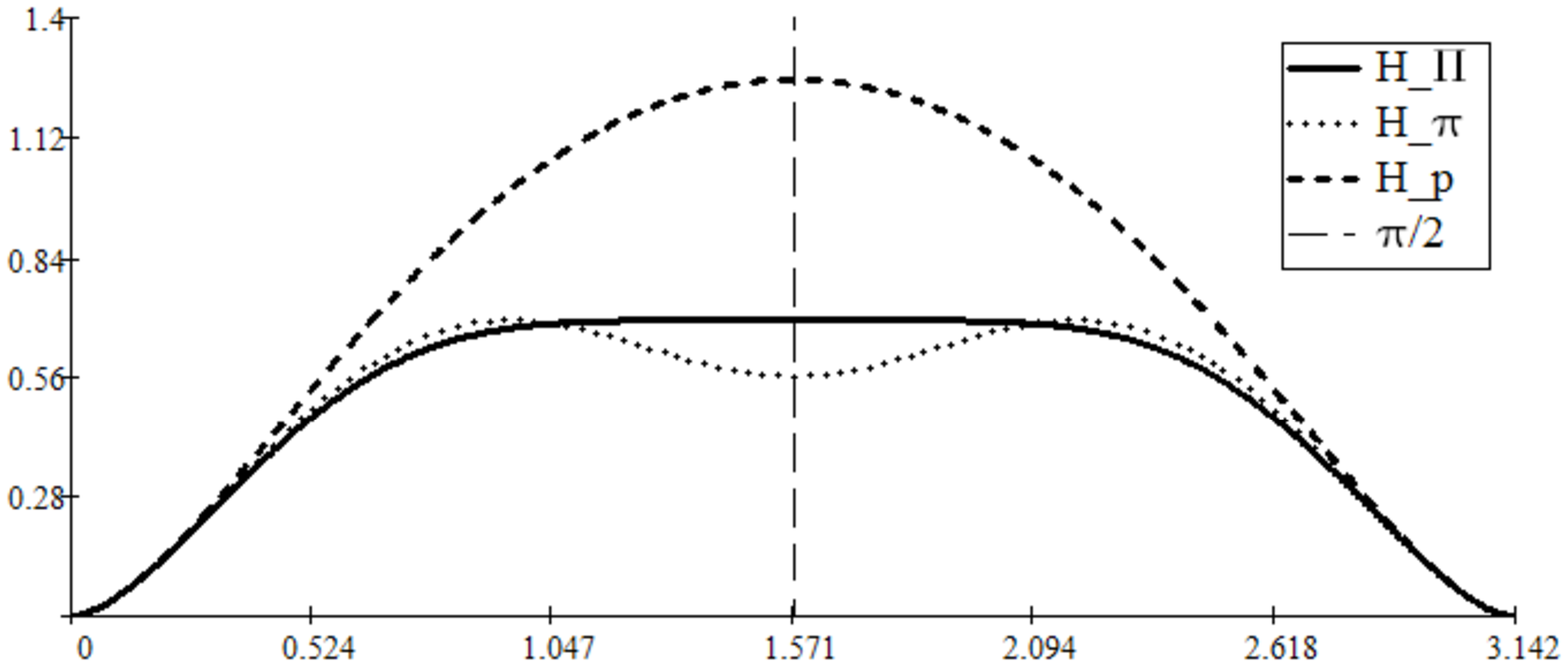}
\vspace{-4mm}
\caption{Entropies $H_{\Pi}(\beta)$, $H_{\pi}(\beta)$, $H_{p}(\beta)$ for probabilities \eqref{13}}
\label{fig:3}
\end{minipage}
\hfill
\begin{minipage}[ht]{0.49\linewidth}
\includegraphics[width=1\linewidth]{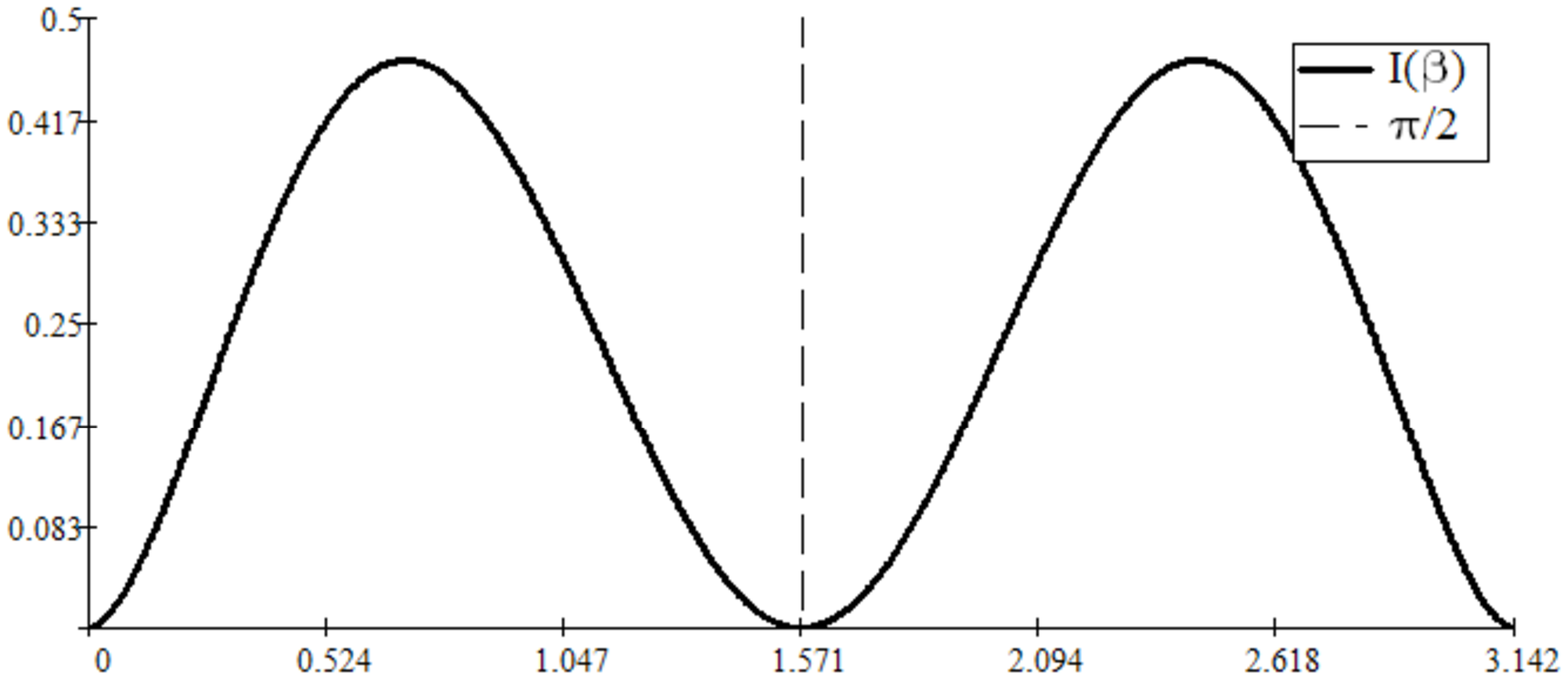}
\vspace{-4mm}
\caption{Information $I(\beta)$ for probabilities \eqref{13}}
\label{fig:4}
\end{minipage}
\end{center}
\end{figure}

\begin{figure}[ht]
\begin{center}
\begin{minipage}[ht]{0.49\linewidth}
\includegraphics[width=1\linewidth]{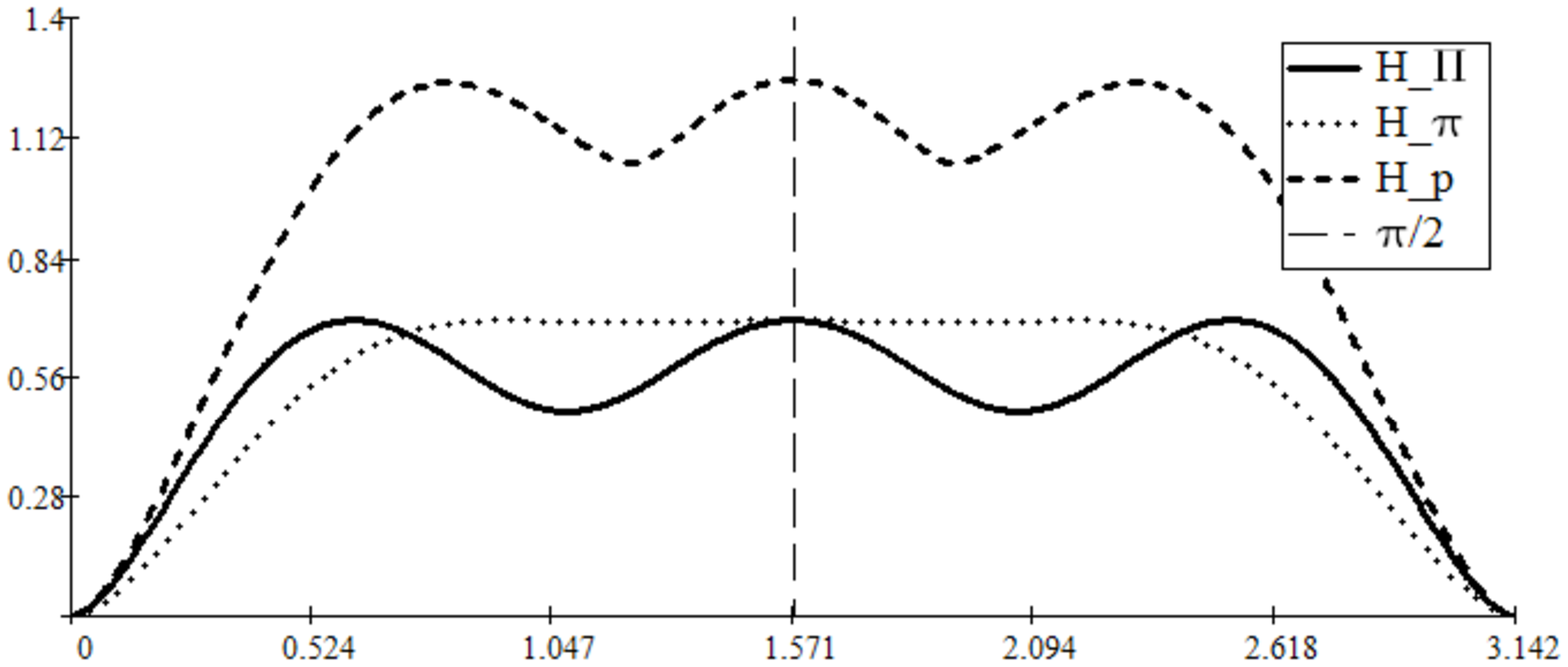}
\vspace{-4mm}
\caption{Entropies $H_{\Pi}(\beta)$, $H_{\pi}(\beta)$, $H_{p}(\beta)$ for probabilities \eqref{20}}
\label{fig:5}
\end{minipage}
\hfill
\begin{minipage}[ht]{0.49\linewidth}
\includegraphics[width=1\linewidth]{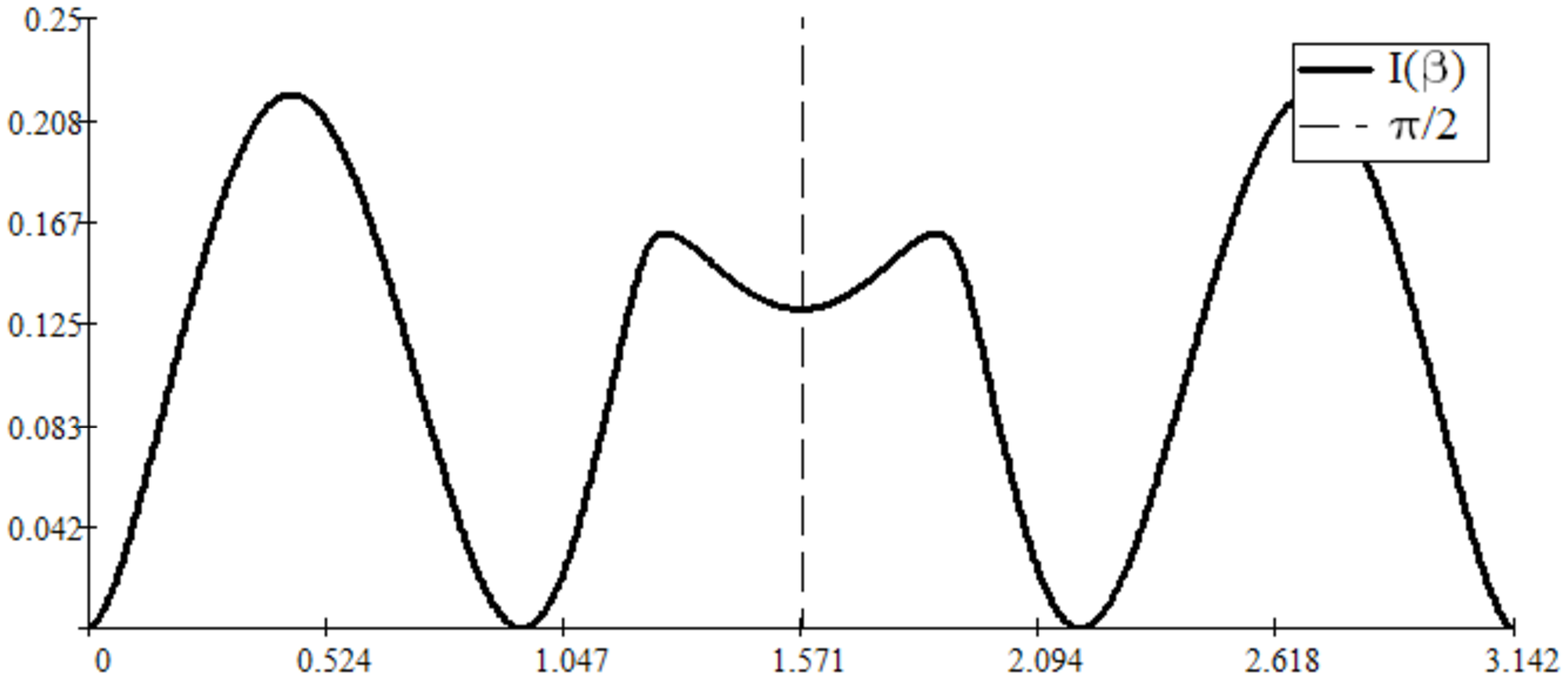}
\vspace{-4mm}
\caption{Information  $I(\beta)$ for probabilities \eqref{20}}
\label{fig:6}
\end{minipage}
\end{center}
\end{figure}

 Let us consider how the permutation of the probabilities \eqref{4} affects on the entropies $H_{\pi}$ and $H_{\Pi}$ \eqref{7}.
The new vector $\widetilde{p}=(\widetilde{p}_1,\widetilde{p}_2,\widetilde{p}_3,\widetilde{p}_4)$ can be determined by
\begin{eqnarray}\label{13}
\widetilde{p}_1\equiv p_{4},\quad \widetilde{p}_2\equiv p_{1},\quad \widetilde{p}_3 \equiv p_{2},\quad\widetilde{ p}_4 \equiv p_{3}.
\end{eqnarray}
Entropies $H_{\Pi}(\beta)$, $H_{\pi}(\beta)$, $H_{p}(\beta)$ and information $I(\beta)$ for the probability vector $\widetilde{p}$ are shown on
Figures \ref{fig:3} and \ref{fig:4}.
Evidently the latter permutations impact only the entropy $H_{\pi}$. The information turns to zero at the point $\beta=\pi/2$.
Substituting \eqref{13} in \eqref{6} and \eqref{7} we can write the following inequality
\begin{eqnarray*}&&-\left(\cos\left(\beta/2\right)^6\widetilde{P}_{0}^{(0,3)}+\frac{\widetilde{P}_{3}^{(-3,0)}}{\sin\left(\beta/2\right)^6}\right)
\ln\left(\cos\left(\beta/2\right)^6\widetilde{P}_{0}^{(0,3)}+\frac{\widetilde{P}_{3}^{(-3,0)}}{\sin\left(\beta/2\right)^6}\right)\\
&-&\left(\frac{\cos\left(\beta/2\right)^4}{3\sin\left(\beta/2\right)^2}
 \widetilde{P}_{1}^{(-1,2)}+\frac{\widetilde{P}_{2}^{(-2,1)}\cos\left(\beta/2\right)^2}{3\sin\left(\beta/2\right)^4}\right)
\ln\left(\frac{\cos\left(\beta/2\right)^4}{3\sin\left(\beta/2\right)^2}
 \widetilde{P}_{1}^{(-1,2)}+\frac{\widetilde{P}_{2}^{(-2,1)}\cos\left(\beta/2\right)^2}{3\sin\left(\beta/2\right)^4}\right)\\
&-&\left(\cos\left(\beta/2\right)^6\widetilde{P}_{0}^{(0,3)}+\frac{\widetilde{P}_{2}^{(-2,1)}\cos\left(\beta/2\right)^2}{3\sin\left(\beta/2\right)^4}\right)
\ln\left(\cos\left(\beta/2\right)^6\widetilde{P}_{0}^{(0,3)}+\frac{\widetilde{P}_{2}^{(-2,1)}\cos\left(\beta/2\right)^2}{3\sin\left(\beta/2\right)^4}\right)\\
&-&\left(\frac{\widetilde{P}_{3}^{(-3,0)}}{\sin\left(\beta/2\right)^6}+\frac{\cos\left(\beta/2\right)^4}{3\sin\left(\beta/2\right)^2}
 \widetilde{P}_{1}^{(-1,2)}\right)
\ln\left(\frac{\widetilde{P}_{3}^{(-3,0)}}{\sin\left(\beta/2\right)^6}+\frac{\cos\left(\beta/2\right)^4}{3\sin\left(\beta/2\right)^2}
 \widetilde{P}_{1}^{(-1,2)}\right)\\
&+&\cos\left(\beta/2\right)^6\widetilde{P}_{0}^{(0,3)}
\ln\left(\cos\left(\beta/2\right)^6\widetilde{P}_{0}^{(0,3)}\right)
+\frac{\cos\left(\beta/2\right)^4}{3\sin\left(\beta/2\right)^2}
 \widetilde{P}_{1}^{(-1,2)}
\ln\left(\frac{\cos\left(\beta/2\right)^4}{3\sin\left(\beta/2\right)^2}
 \widetilde{P}_{1}^{(-1,2)}\right)\\
&+&\frac{\widetilde{P}_{2}^{(-2,1)}\cos\left(\beta/2\right)^2}{3\sin\left(\beta/2\right)^4}
\ln\left(\frac{\widetilde{P}_{2}^{(-2,1)}\cos\left(\beta/2\right)^2}{3\sin\left(\beta/2\right)^4}\right)
+\frac{\widetilde{P}_{3}^{(-3,0)}}{\sin\left(\beta/2\right)^6}
\ln\left(\frac{\widetilde{P}_{3}^{(-3,0)}}{\sin\left(\beta/2\right)^6}\right)
\geq0.
\end{eqnarray*}
Let us select the probability vector $\overrightarrow{p}$ using another combination of $\widetilde{d}_{m',m}^{(j)}(\beta)$. For the fixed  $m=1/2$ we take the elements of the second column in Table \ref{tab:1}. Then the probability vector has the following components
\begin{eqnarray}\label{20}p_1(\beta)&=&\widetilde{d}_{3/2,1/2}^{(j)}(\beta)
=3\cos\left(\beta/2\right)^4\sin\left(\beta/2\right)^2P_{0}^{(1,2)}(\cos\beta)^2
=3\cos\left(\frac{\beta}{2}\right)^4\sin\left(\frac{\beta}{2}\right)^2,\\\nonumber
 p_2(\beta)&=&\widetilde{d}_{1/2,1/2}^{(j)}(\beta)=\frac{(\cos\beta+1)}{2}P_{1}^{(0,1)}(\cos\beta)^2
= \frac{(3\cos\beta-1)^2}{8}(\cos\beta+1),\\\nonumber
p_3(\beta)&=&\widetilde{d}_{-1/2,1/2}^{(j)}(\beta)=\frac{P_{2}^{(-1,0)}(\cos\beta)^2}{\sin\left(\beta/2\right)^2}
= -\frac{(3\cos\beta+1)^2}{8}(\cos\beta-1),\\\nonumber
 p_4(\beta)&=&\widetilde{d}_{-3/2,1/2}^{(j)}(\beta)
 =\frac{3P_{3}^{(-2,-1)}(\cos\beta)^2}{\cos(\beta/2)^2(\cos(\beta/2)^2-1)^2}
=  \frac{3(\cos\beta-1)^2}{8}(\cos\beta+1).
\end{eqnarray}
Substituting \eqref{20} in \eqref{7} we can write the following inequality
\begin{eqnarray*}&&-\left(\frac{\widetilde{P}_{2}^{(-1,0)}}{\sin\left(\beta/2\right)^2}+\frac{3\widetilde{P}_{3}^{(-2,-1)}}{\cos(\beta/2)^2(\cos(\beta/2)^2-1)^2}\right)
\ln\left(\frac{\widetilde{P}_{2}^{(-1,0)}(\cos\beta)^2}{\sin\left(\beta/2\right)^2}+\frac{3\widetilde{P}_{3}^{(-2,-1)}}{\cos(\beta/2)^2(\cos(\beta/2)^2-1)^2}\right)\\
&-&\!\!\!\!\left(\frac{(\cos\beta+1)}{2}\widetilde{P}_{1}^{(0,1)}\!+\!3\cos\left(\frac{\beta}{2}\right)^4\sin\left(\frac{\beta}{2}\right)^2\widetilde{P}_{0}^{(1,2)}\right)
 \!\ln\left(\frac{(\cos\beta+1)}{2}\widetilde{P}_{1}^{(0,1)}\!+\!3\cos\left(\frac{\beta}{2}\right)^4\sin\left(\frac{\beta}{2}\right)^2\widetilde{P}_{0}^{(1,2)}\right)\\
&-&\!\!\!\!\left(\frac{(\cos\beta+1)}{2}\widetilde{P}_{1}^{(0,1)}+\frac{3\widetilde{P}_{3}^{(-2,-1)}}{\cos(\beta/2)^2(\cos(\beta/2)^2-1)^2}\right)
\!\ln\left(\frac{(\cos\beta+1)}{2}\widetilde{P}_{1}^{(0,1)}+\frac{3\widetilde{P}_{3}^{(-2,-1)}}{\cos(\beta/2)^2(\cos(\beta/2)^2-1)^2}\right)\\
&-&\!\left(\frac{\widetilde{P}_{2}^{(-1,0)}}{\sin\left(\beta/2\right)^2}+3\cos\left(\beta/2\right)^4\sin\left(\beta/2\right)^2\widetilde{P}_{0}^{(1,2)}\right)
\ln\left(\frac{\widetilde{P}_{2}^{(-1,0)}}{\sin\left(\beta/2\right)^2}+3\cos\left(\beta/2\right)^4\sin\left(\beta/2\right)^2\widetilde{P}_{0}^{(1,2)}\right)\\
&+&\!\left(3\cos\left(\beta/2\right)^4\sin\left(\beta/2\right)^2\widetilde{P}_{0}^{(1,2)}\right)\ln\left(3\cos\left(\beta/2\right)^4\sin\left(\beta/2\right)^2\widetilde{P}_{0}^{(1,2)}(\cos\beta)^2\right)\\
&+&\!\left(\frac{\widetilde{P}_{2}^{(-1,0)}(\cos\beta)^2}{\sin\left(\beta/2\right)^2}\right)\ln\left(\frac{\widetilde{P}_{2}^{(-1,0)}(\cos\beta)^2}{\sin\left(\beta/2\right)^2}\right)
+\left(\frac{(\cos\beta+1)}{2}\widetilde{P}_{1}^{(0,1)}\right)\ln\left(\frac{(\cos\beta+1)}{2}\widetilde{P}_{1}^{(0,1)}(\cos\beta)^2\right)\\
&+&\!\left(\frac{3\widetilde{P}_{3}^{(-2,-1)}}{\cos(\beta/2)^2(\cos(\beta/2)^2-1)^2}\right)\ln\left(\frac{3\widetilde{P}_{3}^{(-2,-1)}}{\cos(\beta/2)^2(\cos(\beta/2)^2-1)^2}\right)\geq0.
\end{eqnarray*}
Entropies $H_{\Pi}(\beta)$, $H_{\pi}(\beta)$, $H_{p}(\beta)$ and information $I(\beta)$ for probability vector \eqref{20} are shown on
Figures \ref{fig:5} and \ref{fig:6}. Evidently, they differ from the entropies and information constructed by the polynomials based on the first column of Table \ref{tab:1}.
\par It is also interesting to see how the permutation of the components of probability vector $\overrightarrow{p}$
will change the information. To this end we do the same procedure as in \eqref{13}.
\par
\begin{figure}[ht]
\begin{center}
\begin{minipage}[ht]{0.49\linewidth}
\includegraphics[width=1\linewidth]{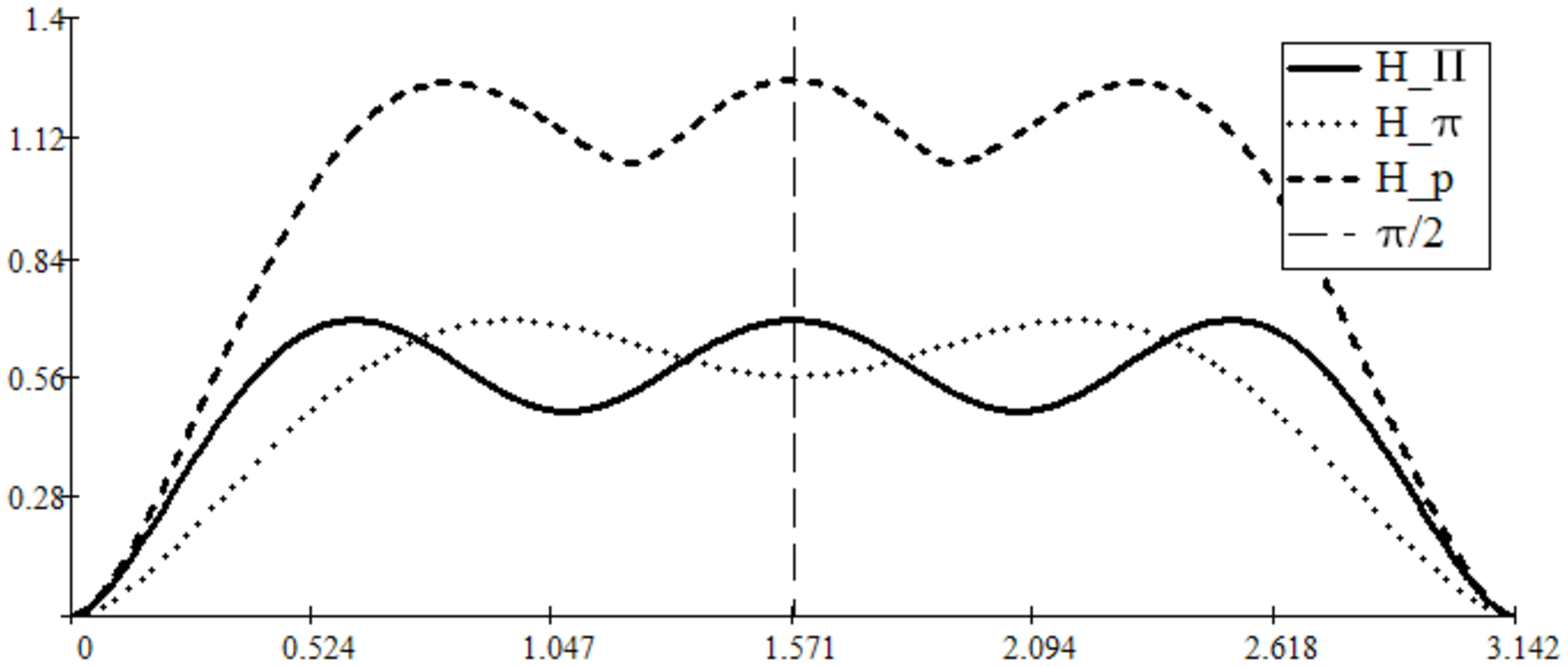}
\vspace{-4mm}
\caption{Entropies $H_{\Pi}(\beta)$, $H_{\pi}(\beta)$, $H_{p}(\beta)$ for probabilities \eqref{21}}
\label{fig:7}
\end{minipage}
\hfill
\begin{minipage}[ht]{0.49\linewidth}
\includegraphics[width=1\linewidth]{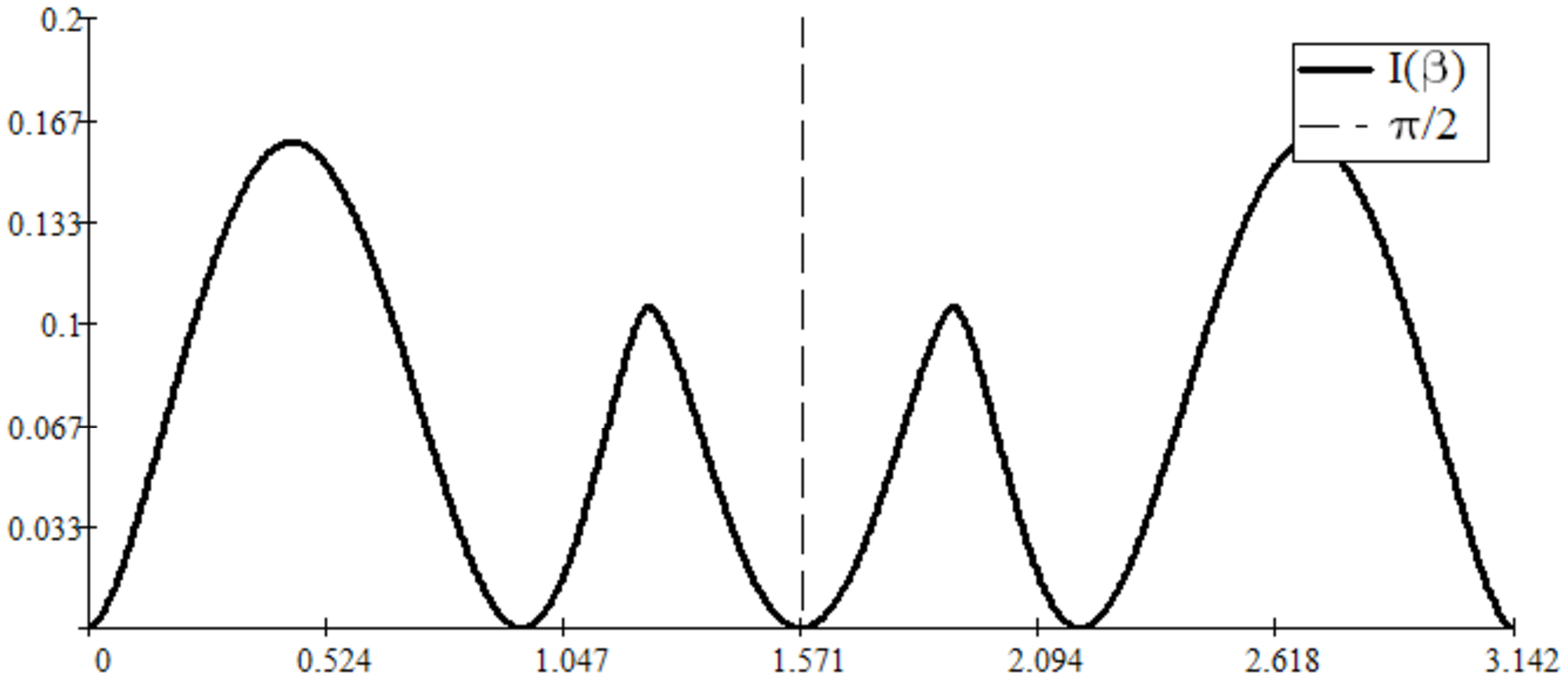}
\vspace{-4mm}
\caption{Information  $I(\beta)$ for probabilities \eqref{21}}
\label{fig:8}
\end{minipage}
\end{center}
\end{figure}

 The new vector $\widehat{p}=(\widehat{p}_1,\widehat{p}_2,\widehat{p}_3,\widehat{p}_4)$ can be formed by
\begin{eqnarray}\label{21}
\widehat{p}_1\equiv p_{4},\quad \widehat{p}_2\equiv p_{1},\quad \widehat{p}_3 \equiv p_{2},\quad\widehat{p}_4 \equiv p_{3}.
\end{eqnarray}
 Thus, substituting \eqref{21} in \eqref{6} and \eqref{7} new entropies and information can be obtained. The latter are shown in Figures  \ref{fig:7} and \ref{fig:8}. In contrast to Figure \ref{fig:6}, the information for permuted vector $\widehat{p}$ equals to zero at point $\beta=\pi/2$.
\par
\begin{figure}[ht]
\begin{center}
\begin{minipage}[ht]{0.49\linewidth}
\includegraphics[width=1\linewidth]{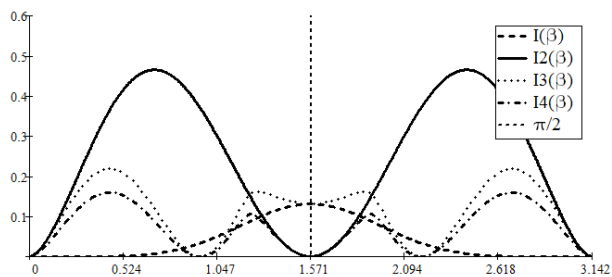}
\vspace{-4mm}
\caption{Information of four probability vectors}
\label{fig:9}
\end{minipage}
\end{center}
\end{figure}

 Finally we summarize all results of Figures \ref{fig:2}, \ref{fig:4}, \ref{fig:6}, \ref{fig:8} in Figure \ref{fig:9}. Obviously, for systems \eqref{4} and \eqref{20} the information is non zero for $\beta=\pi/2$ and for permutated vectors the information is zero.
\par From \eqref{12} it is easy to verify that the latter two examples cover all possible probabilities.
Other combinations chosen from Table \ref{tab:1} determine their permutations. However, Jacobi polynomials corresponding to them are of course not the same. It allows us to obtain many different inequalities of the form \eqref{17}.
\section{Case of the $N$-component probability  vector}
\pst
In this section, we shall extend the latter theory on the case when the probability vector has $N$ components, i.e.  $\overrightarrow{p}=(p_1,p_2,\ldots,p_N)$. If $N$ is an even number than we can introduce the map of indices similar to \eqref{14}, namely it holds
\begin{eqnarray*}
1&\Leftrightarrow& 11,\quad 2\Leftrightarrow 12,\quad 3 \Leftrightarrow 13,\quad\ldots,\quad\frac{N}{2} \Leftrightarrow 1\frac{N}{2},\\
\frac{N}{2}+1&\Leftrightarrow& 21,\quad \frac{N}{2}+2\Leftrightarrow 22,\quad \frac{N}{2}+3 \Leftrightarrow 23,\quad\ldots,\quad N \Leftrightarrow 2\frac{N}{2}.
\end{eqnarray*}
Hence the probabilities are given in the form of the matrix $(p_{il})$, $i=1,2$, $l=1,2,\ldots,\frac{N}{2}$ with components
\begin{eqnarray}\label{23}
p_1&\Leftrightarrow& p_{11},\quad p_2\Leftrightarrow p_{12},\quad p_3 \Leftrightarrow p_{13},\quad\ldots,\quad p_{\frac{N}{2}} \Leftrightarrow p_{1\frac{N}{2}},\\\nonumber
p_{\frac{N}{2}+1}&\Leftrightarrow& p_{21},\quad p_{\frac{N}{2}+2}\Leftrightarrow p_{22},\quad p_{\frac{N}{2}+3} \Leftrightarrow p_{23},\quad\ldots,\quad p_{N} \Leftrightarrow p_{2\frac{N}{2}}.
\end{eqnarray}
If $N$ is an odd number than we assign a zero vector $p_{N+1}=0$ to the $N$-component vector $\overrightarrow{p}$. Then we get $(N+1)$-component vector $\overrightarrow{p}=(p_1,p_2,\ldots,p_N,p_{N+1})$.
Thus the insertable map of the indices is
\begin{eqnarray*}
1&\Leftrightarrow& 11,\quad 2\Leftrightarrow 12,\quad 3 \Leftrightarrow 13,\quad\ldots,\quad\frac{N+1}{2} \Leftrightarrow 1\frac{N+1}{2},\\
\frac{N+1}{2}+1&\Leftrightarrow& 21,\quad \frac{N+1}{2}+2\Leftrightarrow 22,\quad \frac{N+1}{2}+3 \Leftrightarrow 23,\quad\ldots,\quad N+1 \Leftrightarrow 2\frac{N+1}{2}.
\end{eqnarray*}
and probabilities are given in the form of the matrix $(p_{il})$, $i=1,2$, $l=1,2,\ldots,\frac{N+1}{2}$ with components
\begin{eqnarray}\label{24}
p_1&\Leftrightarrow& p_{11},\quad p_2\Leftrightarrow p_{12},\quad p_3 \Leftrightarrow p_{13},\quad\ldots,\quad p_{\frac{N+1}{2}} \Leftrightarrow p_{1\frac{N+1}{2}},\\\nonumber
p_{\frac{N+1}{2}+1}&\Leftrightarrow& p_{21},\quad p_{\frac{N+1}{2}+2}\Leftrightarrow p_{22},\quad p_{\frac{N+1}{2}+3} \Leftrightarrow p_{23},\quad\ldots,\quad p_{N+1} \Leftrightarrow p_{2\frac{N+1}{2}}.
\end{eqnarray}
Similarly  to \eqref{15} we can write marginal probability distributions determined by the joint probability distribution. For
an odd $N$ these are the following
\begin{eqnarray}\label{16}\pi_i&=&\sum\limits_{l=1}^{\frac{N+1}{2}}p_{il}=p_{i1}+p_{i2}+p_{i3}+\cdots+p_{i\frac{N+1}{2}},\quad i=1,2\\ \nonumber
\Pi_l&=&\sum\limits_{i=1}^2p_{il}=p_{1l}+p_{2l},\quad l=1,2,\ldots,\frac{N+1}{2}.
\end{eqnarray}
For \eqref{16} it is possible to obtain inequalities similar to \eqref{17}. To this end, we define Shannon entropies by
 \begin{eqnarray}\label{18}H_{p}&=&-\sum\limits_{t=1}^{N+1}p_t\ln p_t,\\\nonumber
H_{\pi_{1}}&=&-\sum\limits_{t=1}^{\frac{N+1}{2}}p_t\ln\left(\sum\limits_{t=1}^{\frac{N+1}{2}}p_t\right),\quad
H_{\pi_{2}}=-\sum\limits_{t=\frac{N+1}{2}+1}^{N+1}p_t\ln\left(\sum\limits_{t=\frac{N+1}{2}+1}^{N+1}p_t\right),\\\nonumber
H_{\Pi_{l}}&=&-\left(p_l+p_{\frac{N+1}{2}+l}\right)\ln\left(p_l+p_{\frac{N+1}{2}+l}\right),\quad l=1,2,\ldots,\frac{N+1}{2}.
\end{eqnarray}
Then the Shannon information that is similar to \eqref{7} is based on the entropies \eqref{18}
\begin{eqnarray}\label{19}I(\beta)_{tl}&=&H_{\pi_{t}}(\beta)+H_{\Pi_{l}}(\beta)-H_{p}(\beta), \quad t=1,2\quad l=1,2,\ldots,\frac{N+1}{2} .
\end{eqnarray}
Obviously the inequality $I(\beta)\geq 0$ for all the angles $\beta$ is valid.
\par The spin projections  $m$ and $m'$ can take $2j+1=N$ values for the spin $j$. If $j$ is a fractional number ($N$ is even), than projections can be $-j,-j+1,\ldots,j-1,j$. Then the components of the probability  $\overrightarrow{p}$  are given by \eqref{23}.
On the other hand, if $j$ is an integer number ($N$ is odd), than spin projections can be $-j,-j+1,\ldots,0,\ldots,j-1,j$. Then the components of the probability  vector are given by \eqref{24}.
\par One can take the polynomials $\Big|d_{m',m}^{(j)}(\beta)\Big|^2$ as the probabilities for a fixed $m'$ and for all $m$.
One of the possible choices for an odd $N$ is the following
\begin{eqnarray}\label{27}p_1&=&\Big|d_{m',j}^{(j)}(\beta)\Big|^2, p_2=\Big|d_{m',j-1}^{(j)}(\beta)\Big|^2,\ldots, p_{\frac{N+1}{2}}=\Big|d_{m',0}^{(j)}(\beta)\Big|^2,
\ldots, p_{N}=\Big|d_{m',-j}^{(j)}(\beta)\Big|^2, p_{N+1}=0.
\end{eqnarray}
In this notations inequality \eqref{19} for $t=1$ and $l=\frac{N+1}{2}$ is given by
\begin{eqnarray}\label{25}-\sum\limits_{m=0}^{j}\widetilde{d}_{m',m}^{(j)}(\beta)\ln\left(\sum\limits_{m=0}^{j}\widetilde{d}_{m',m}^{(j)}(\beta)\right)
-\widetilde{d}_{m',0}^{(j)}(\beta)\ln\left(\widetilde{d}_{m',0}^{(j)}(\beta)\right)+\sum\limits_{m=-j}^{j}\widetilde{d}_{m',m}^{(j)}(\beta)
\ln\left(\sum\limits_{m=-j}^{j}\widetilde{d}_{m',m}^{(j)}(\beta)\right)\geq0
\end{eqnarray}
or for $t=2$ and $l=1$ it looks like
\begin{eqnarray}\label{26}&-&\sum\limits_{m=-j}^{-1}\widetilde{d}_{m',m}^{(j)}(\beta)\ln\left(\sum\limits_{m=-j}^{-1}\widetilde{d}_{m',m}^{(j)}(\beta)\right)
-\left(\widetilde{d}_{m',j}^{(j)}(\beta)+\widetilde{d}_{m',-1}^{(j)}(\beta)\right)\ln\left(\widetilde{d}_{m',j}^{(j)}(\beta)+\widetilde{d}_{m',-1}^{(j)}(\beta)\right)\\\nonumber
&+&\sum\limits_{m=-j}^{j}\widetilde{d}_{m',m}^{(j)}(\beta)
\ln\left(\sum\limits_{m=-j}^{j}\widetilde{d}_{m',m}^{(j)}(\beta)\right)\geq0.
\end{eqnarray}
Our aim is to represent \eqref{25} and \eqref{26} by Jacobi polynomials. To this end, we introduce the new notation in \eqref{1}
\begin{eqnarray*}
\widetilde{d}_{m',m}^{(j)}(\beta)&=&\Big|d_{m',m}^{(j)}(\beta)\Big|^2=G_{m',m}^{j}(\beta)\cdot \widetilde{P}_{j-m'}^{(m'-m,m'+m)},
\end{eqnarray*}
where
\begin{eqnarray*}
G_{m',m}^{j}(\beta)&=&\frac{(j+m')!(j-m')!}{(j+m)!(j-m)!}\left(\cos\left(\frac{\beta}{2}\right)^{m'+m}\sin\left(\frac{\beta}{2}\right)^{m'-m}\right)^2.
\end{eqnarray*}
Hence inequalities \eqref{25} and \eqref{26} can be rewritten in the new terms as
\begin{eqnarray*}&-&\sum\limits_{m=0}^{j}G_{m',m}^{j}(\beta)\widetilde{P}_{j-m'}^{(m'-m,m'+m)}\ln\left(\sum\limits_{m=0}^{j}G_{m',m}^{j}(\beta) \widetilde{P}_{j-m'}^{(m'-m,m'+m)}\right)\\
&-&G_{m',0}^{j}(\beta)\widetilde{P}_{j-m'}^{(m',m')}\ln\left(G_{m',0}^{j}(\beta)\widetilde{P}_{j-m'}^{(m',m')}\right)\\
&+&\sum\limits_{m=-j}^{j}G_{m',m}^{j}(\beta)\widetilde{P}_{j-m'}^{(m'-m,m'+m)}
\ln\left(\sum\limits_{m=-j}^{j}G_{m',m}^{j}(\beta)\widetilde{P}_{j-m'}^{(m'-m,m'+m)}\right)\geq0,
\end{eqnarray*}
\begin{eqnarray*}&-&\sum\limits_{m=-j}^{-1}G_{m',m}^{j}(\beta)\widetilde{P}_{j-m'}^{(m'-m,m'+m)}\ln\left(\sum\limits_{m=-j}^{-1}G_{m',m}^{j}(\beta) \widetilde{P}_{j-m'}^{(m'-m,m'+m)}\right)\\
&-&\left(G_{m',j}^{j}(\beta)\widetilde{P}_{j-m'}^{(m'-j,m'+j)}+G_{m',-1}^{j}(\beta)\cdot \widetilde{P}_{j-m'}^{(m'+1,m'-1)}\right)\cdot\\
&\cdot&\ln\left(G_{m',j}^{j}(\beta\widetilde{P}_{j-m'}^{(m'-j,m'+j)}+G_{m',-1}^{j}(\beta) \widetilde{P}_{j-m'}^{(m'+1,m'-1)}\right)\\\nonumber
&+&\sum\limits_{m=-j}^{j}G_{m',m}^{j}(\beta)\widetilde{P}_{j-m'}^{(m'-m,m'+m)}
\ln\left(\sum\limits_{m=-j}^{j}G_{m',m}^{j}(\beta)\widetilde{P}_{j-m'}^{(m'-m,m'+m)}\right)\geq0.
\end{eqnarray*}
Let us consider a special case when $j=c$, where $c$  is an integer number and $m'=0$. Then \eqref{1} can be rewritten as
\begin{eqnarray}\label{22}
\Big|d_{0,m}^{(c)}(\beta)\Big|^2&=&\Big|d_{m,0}^{(c)}(\beta)\Big|^2=\frac{(c-m)!}{(c+m)!}\left(P_{c}^{m}(\cos\beta)\right)^2,
\end{eqnarray}
where $P_{c}^{m}(\cos\beta)$ are the  Legendre polynomials \cite{Landau}.
\par Hence \eqref{25} for this special case is determined by
\begin{eqnarray*}-\!\!\sum\limits_{m=0}^{j}\frac{(c-m)!}{(c+m)!}\widetilde{P}_{c}^{m}\ln\left(\sum\limits_{m=0}^{j}\frac{(c-m)!}{(c+m)!}\widetilde{P}_{c}^{m}\right)
-\widetilde{P}_{c}^{0}\ln\left(\widetilde{P}_{c}^{0}\right)+\sum\limits_{m=-j}^{j}\frac{(c-m)!}{(c+m)!}\widetilde{P}_{c}^{m}
\ln\left(\sum\limits_{m=-j}^{j}\frac{(c-m)!}{(c+m)!}\widetilde{P}_{c}^{m}\right)\geq0
\end{eqnarray*}
\begin{figure}[ht]
\begin{center}
\begin{minipage}[ht]{0.49\linewidth}
\includegraphics[width=1\linewidth]{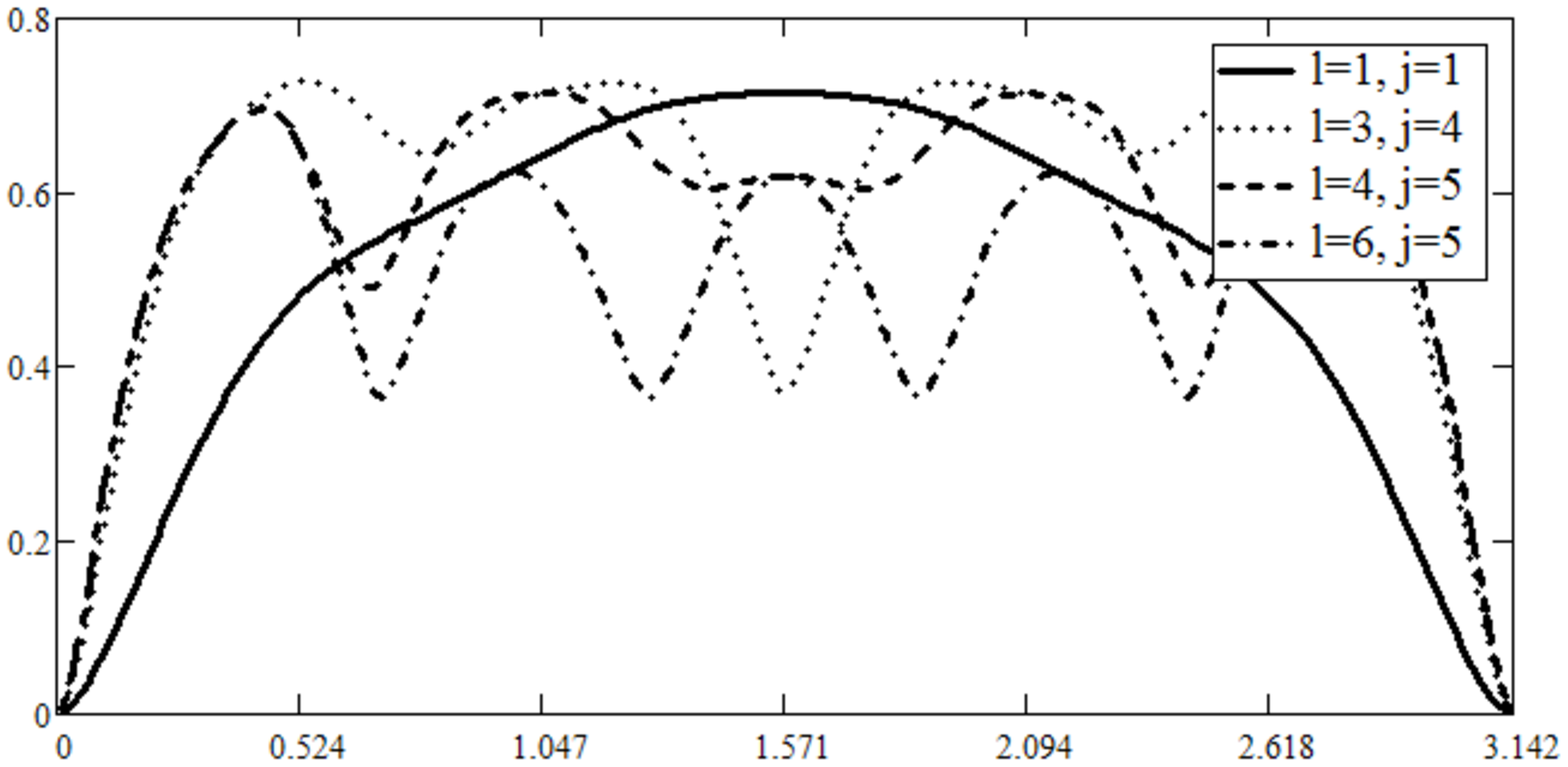}
\vspace{-4mm}
\caption{Shannon information \eqref{19} for probabilities \eqref{27}, $t=1$}
\label{fig:10}
\end{minipage}
\hfill
\begin{minipage}[ht]{0.49\linewidth}
\includegraphics[width=1\linewidth]{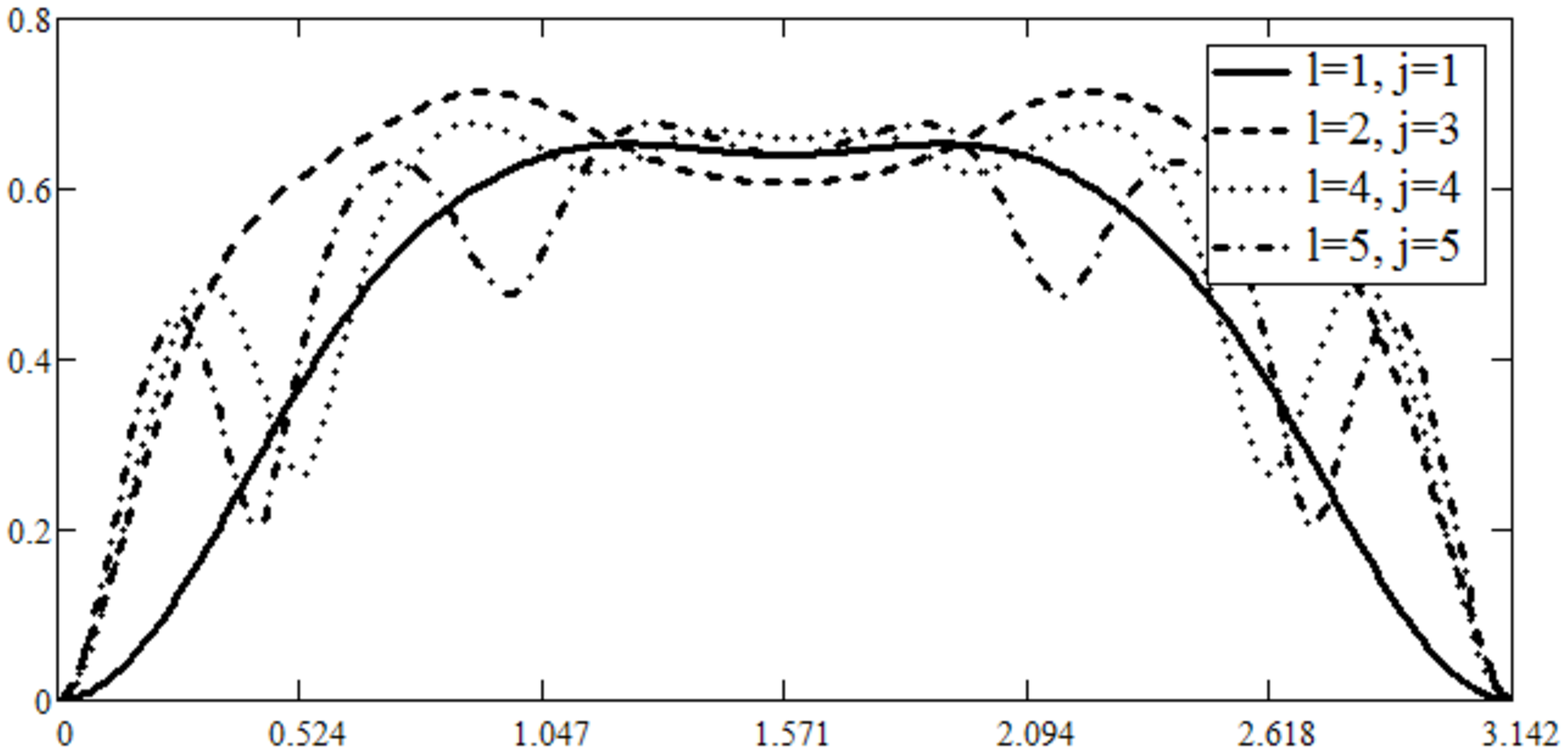}
\vspace{-4mm}
\caption{Shannon information \eqref{19} for probabilities \eqref{27}, $t=2$}
\label{fig:11}
\end{minipage}
\end{center}
\end{figure}
\\and \eqref{26} is rewritten by
\begin{eqnarray*}&-&\sum\limits_{m=-j}^{-1}\frac{(c-m)!}{(c+m)!}\widetilde{P}_{c}^{m}\ln\left(\sum\limits_{m=-j}^{-1}\frac{(c-m)!}{(c+m)!}\widetilde{P}_{c}^{m}\right)
-\frac{(c+1)!}{(c-1)!}\widetilde{P}_{c}^{-1}\ln\left(\frac{(c+1)!}{(c-1)!}\widetilde{P}_{c}^{-1}\right)\\\nonumber
&+&\sum\limits_{m=-j}^{j}\frac{(c-m)!}{(c+m)!}\widetilde{P}_{c}^{m}
\ln\left(\sum\limits_{m=-j}^{j}\frac{(c-m)!}{(c+m)!}\widetilde{P}_{c}^{m}\right)\geq0.
\end{eqnarray*}
Here we use the notation $\left(P_{c}^{m}(\cos\beta)\right)^2\equiv\widetilde{P}_{c}^{m}$ for Legendre polynomials.

\par Information \eqref{19} is shown in Figures \ref{fig:10} and \ref{fig:11} for entropies \eqref{18} and various parameters $t$, $l$ and spins $j$.
\section{Summary}
\pst
To conclude we point out the main results of the work. Considering the matrix elements of the unitary irreducible
representations of the group $SU(2)$ and applying known subadditivity condition for joint probability distributions constructed from these matrix elements we obtained new inequalities for the Jacobi and Legendre polynomials. The inequalities correspond to entropic inequalities for Shannon entropies
of bipartite classical systems. The results are shown in detail on the example of spin $j=3/2$, where the Shannon  information of the bipartite system is expressed in terms of the polynomials. The general approach to get analogous information and entropic inequalities for the arbitrary spins $j$ is formulated.

\section*{Acknowledgments}
\pst
L. A. M. acknowledges the financial support provided within the Russian Foundation for Basic Research, grant 13-08-00744 A.


\begin{thebibliography}{99}
\bibitem{Vil}
 N. Ja Vilenkin, A. U. Klimyk, {\it Representation of Lie Groups and Special Functions: Recent Advances (Mathematics and Its Applications)}, Springer (1994).

\bibitem{Ibort}
A. Ibort,  V. I. Man'ko, G. Marmo, A. Simoni, F. Ventriglia,
{\it An introduction to the tomographic picture of quantum mechanics}, Physica Scripta \textbf{79(6)} (2009).

\bibitem{Shannon}C. E. Shannon,  {\it A Mathematical Theory of Communication}, Bell System Technical Journal, \textbf{27}, 379 (1948).

\bibitem{Nelsen}R. B. Nelsen, \it{An Introduction to Copulas}, Springer (2006).

\bibitem{Lieb}E. H. Lieb, M. B. Ruskai,  {\sl J. Math. Phys.}, \textbf{14}, (1938).

\bibitem{Chernega}V. N. Chernega and V. I. Man'ko, {\sl J. Russ. Laser Res.}, \textbf{29}, 505 (2008).

\bibitem{Mendes}M. A. Man'ko, V. I. Man'ko, and R. Vilela Mendes, {\sl J. Russ. Laser Res.}, \textbf{27}, 507 (2006).

\bibitem{Landau}L. D. Landau, E.M. Lifshitz,
{\it Quantum Mechanics: Non-Relativistic Theory}, Butterworth-Heinemann (1977).






\end{thebibliography}
\end{document}